\documentclass[prd,aps, tightenlines, preprintnumbers, showpacs, nofootinbib,superscriptaddress,notitlepage,11pt]{revtex4-1}

\usepackage{epsfig}
\usepackage{amsmath,amsthm,amssymb}
\usepackage{mathtools}
\usepackage{siunitx}

\usepackage[usenames,dvipsnames]{color}

\newcommand{\beq}{\begin{eqnarray}}
\newcommand{\eeq}{\end{eqnarray}}

\newcommand{\nn}{\nonumber \\}

\newif\iffigure
\figuretrue

\begin{document}
\preprint{YITP-18-08}

\title{On the small-$x$ behavior of the orbital angular momentum distributions\\ in QCD }

\author{Yoshitaka Hatta}
\affiliation{Yukawa Institute for Theoretical Physics, Kyoto University, Kyoto 606-8502, Japan}

\author{Dong-Jing Yang}
\affiliation{Department of Physics, National Taiwan Normal University, Taipei 10610, Taiwan, Republic of China}


\begin{abstract}
We present the numerical solution of the leading order QCD evolution equation for the orbital angular momentum distributions of quarks and gluons and discuss its implications for the nucleon spin sum rule. We observe that at small-$x$, the gluon helicity and orbital angular momentum  distributions are roughly of the same magnitude but with opposite signs, indicating a significant cancellation between them. A similar cancellation occurs also in the quark sector. We explain analytically the reason for this cancellation. 
\end{abstract}
\maketitle

\section{Introduction}

Over the past decades, tremendous effort has been poured into determining the partonic helicity contributions to the nucleon spin. It has been known for quite some time that quarks' helicity $\Delta \Sigma$ accounts  for only a quarter of  the nucleon spin. A recent NLO global QCD analysis has found a nonzero contribution from the helicity of gluons $\Delta G$ \cite{deFlorian:2014yva}. When combined, these two contributions come closer to, but still fall short of the expected value of $\frac{1}{2}$. One might expect that the remaining discrepancy could be resolved by a precise future determination of the gluon helicity distribution $\Delta G(x)$ in the small-$x$ region where the current theoretical uncertainties are overwhelmingly large.     

However, {\it a priori} there is no reason to expect that the nucleon spin entirely originates from partons' helicity. 
As is clear from the Jaffe-Manohar sum rule \cite{Jaffe:1989jz}, 
\beq
\frac{1}{2}=\frac{1}{2}\Delta \Sigma + \Delta G+L_q+L_g. \label{1}
\eeq
the resolution of the spin puzzle cannot be complete without a full understanding of the orbital angular momentum of quarks $L_{q}$ and gluons $L_g$ \cite{Bashinsky:1998if,Hatta:2011ku}. Unfortunately, at the moment very little is known about the actual value of $L_{q,g}$, and the community still has a long way to go in extracting them from experiments.  
   The recent proposals of observables for $L_{q,g}$ \cite{Hatta:2016aoc,Ji:2016jgn,Bhattacharya:2017bvs} as well as the first lattice QCD computation of $L_q$ \cite{Engelhardt:2017miy} are particularly encouraging in this direction.


In this paper, we investigate the QCD evolution of the orbital angular momentum. The four terms in (\ref{1}) actually depend on  the renormalization scale $Q^2$ \cite{Ji:1995cu}.  Moreover, they  can be written as the integral over Bjorken-$x$ of the corresponding partonic distributions. For $\Delta\Sigma$ and $\Delta G$, these are the usual polarized parton distributions $\Delta\Sigma(x)$ and $\Delta G(x)$. A less known fact is that the $x$-distributions for $L_{q,g}$ can also be defined \cite{Harindranath:1998ve,Hagler:1998kg,Bashinsky:1998if,Hatta:2012cs,Ji:2012ba}
\beq
L_{q,g}(Q^2)=\int_0^1dx L_{q,g}(x,Q^2).
\eeq 
The $Q^2$-evolution of $L_{q,g}(x,Q^2)$ has been previously studied in \cite{Martin:1998fe} by solving the renormalization group  equation for  the moments $L_{q,g}(j,Q^2)=\int_0^1 dx x^{j-1}L_{q,g}(x,Q^2)$ and performing the inverse Mellin transformation. Now that we know the fully gauge invariant definitions of $L_{q,g}(x,Q^2)$ and their detailed twist structure \cite{Hatta:2012cs}, we think it is timely and worthwhile to revisit this problem, by numerically solving the DGLAP-like  evolution equation  for $L_{q,g}(x,Q^2)$ directly in the $x$-space. We shall be particularly interested in the small-$x$ behavior of $L_{q,g}(x)$ which was not a focus of interest in \cite{Martin:1998fe}, but is phenomenologically important in view of the recent controversy over the large uncertainties in $\Delta G$.
Indeed, we  find that there is a significant cancellation between the helicity and orbital angular momentum distributions at small-$x$ both in the quark and gluon sectors. We explain analytically that such a cancellation is a robust feature of the evolution equation.

In this work, we restrict ourselves to the $Q^2$-evolution equation, and do not discuss the evolution equation in $x$. The latter requires an intricate resummation of double logarithms $\alpha_s \ln^2 1/x$ which has recently enjoyed renewed interest for the helicity distributions 
\cite{Bartels:1996wc,Bartels:1995iu,Kovchegov:2015pbl,Kovchegov:2016zex,Kovchegov:2017lsr}, but not yet for the orbital angular momentum distributions. While we push our numerical simulation down to very small values of $x$, one has to keep in mind that at some point the DGLAP equation  breaks down, and should be smoothly taken over by the small-$x$ equation. Where and how exactly this transition occurs is presently not understood.

\section{Evolution equation}

Let us express the four terms in (\ref{1}) as the first moment in Bjorken-$x$ of the corresponding partonic distribution functions
\beq
\Delta\Sigma(Q^2) &=& \sum_{f}\int_0^1dx (\Delta q_f(x,Q^2)+\Delta \bar{q}_f(x,Q^2)), \nn
 \quad \Delta G(Q^2)&=&\int_0^1dx \Delta G(x,Q^2), \nn
  L_{q}(Q^2)&=&\sum_f \int_0^1dx (L_{f}(x,Q^2)+\bar{L}_f(x,Q^2)), \nn 
L_g(Q^2)&=&\int_0^1 dx L_g(x,Q^2), \label{int}
\eeq
 where $f$ is the flavor index. The polarized quark $\Delta q_f(x)$, antiquark $\Delta\bar{q}_f(x)$ and gluon $\Delta G(x)$ helicity distributions are standard, whereas the orbital angular momentum distributions for quarks $L_f(x)$, antiquarks $\bar{L}_f(x)$ and gluons $L_g(x)$ are perhaps less familiar.
Their operator definitions in   the light-cone gauge ($A^+=0$) were first introduced in \cite{Harindranath:1998ve} (see also \cite{Hagler:1998kg}). The fully gauge invariant definitions of $L_{q,g}(x)$ in QCD have been obtained in \cite{Hatta:2012cs}\footnote{
In Ref.\cite{Hatta:2012cs}, $L_{q,g}(x)$ have been defined such that 
\beq
 \int_{-1}^1dx L^{{\rm there}}_q(x)=L_q\,, \qquad \int_{-1}^1 dx L^{{\rm there}}_g(x) =L_g\,.
\eeq
 These are related to the present convention as (Note that $L_g(x)$ is an even function of $x$.)
\beq
\begin{cases} \sum_f L^{{\rm here}}_f(x) =L^{{\rm there}}_q(x), \\ \sum_f \bar{L}^{{\rm here}}_f(x) =L^{{\rm there}}_q(-x), \end{cases}
   L_g^{{\rm here}}(x)=2L_g^{{\rm there}}(x). \qquad  (x>0)
\eeq
} (see, also, \cite{Bashinsky:1998if,Ji:2012ba})
 where it has been shown that $L_{q}(x)$ and $L_g(x)$ are not usual twist-two distributions, but have both the twist-two (`Wandzura-Wilczek') and genuine twist-three components, so like the $g_2(x)$ structure function for the transversely polarized nucleon.

In this paper we only consider the singlet distributions 
\beq
\Delta \Sigma(x,Q^2)&\equiv& \sum_f (\Delta q_f(x,Q^2)+\Delta \bar{q}_f(x,Q^2)) , \nn 
L_q(x,Q^2)&\equiv& \sum_f  (L_f(x,Q^2) + \bar{L}_f(x,Q^2)).
\eeq
The four distributions $\Delta\Sigma(x,Q^2)$, $\Delta G(x.Q^2)$. $L_{q,g}(x,Q^2)$ satisfy the renormalization group equation in $Q^2$. For $\Delta\Sigma(x)$ and $\Delta G(x)$, this is the well-known DGLAP equation which reads, to leading order, 
\beq
\frac{d}{d\ln Q^2} \left(\begin{matrix} \Delta \Sigma(x) \\ \Delta G(x) \end{matrix}\right)= \frac{\alpha_s(Q^2)}{2\pi} \int_x^1 \frac{dz}{z} \left(\begin{matrix} \Delta P_{qq}(z) &  \Delta P_{qg}(z) \\  \Delta P_{gq}(z) & \Delta P_{gg}(z) \end{matrix}\right) \left(\begin{matrix}   \Delta \Sigma(x/z) \\ \Delta G(x/z) \end{matrix}\right)\,, \label{d1}
\eeq
where 
\beq
&&\Delta P_{qq}(z)=C_F \left(\frac{1+z^2}{(1-z)_+} + \frac{3}{2}\delta(1-z) \right)\,, \\
&&\Delta P_{qg}(z)= n_f(2z-1) \,, \\
&&\Delta P_{gq}(z)= C_F (2-z)\,, \\
&&\Delta P_{gg}(z)=6 \left(\frac{1}{(1-z)_+}-2z+1\right) +  \frac{\beta_0}{2}\delta(z-1)\,,
\eeq
with $C_F=\frac{N_c^2-1}{2N_c}=\frac{4}{3}$, $n_f$ being the number of flavors and $\beta_0=11-\frac{2n_f}{3}$.  

 The corresponding equation for $L_{q,g}(x,Q^2)$ has been implicitly derived in \cite{Hagler:1998kg} to one-loop and explicitly written down in \cite{Hatta:2016aoc}. 
 Because $L_{q,g}(x)$ have a twist-two component, they mix with $\Delta q(x)$ and $\Delta G(x)$ under renormalization as
\beq
\frac{d}{d\ln Q^2} \left(\begin{matrix} L_q(x) \\ L_g(x) \end{matrix}\right)= \frac{\alpha_s}{2\pi} \int_x^1 \frac{dz}{z} \left(\begin{matrix} \hat{P}_{qq}(z)  & \hat{P}_{qg}(z) & \Delta \hat{P}_{qq}(z) & \Delta \hat{P}_{qg}(z) \\ \hat{P}_{gq}(z) & \hat{P}_{gg}(z) & \Delta \hat{P}_{gq}(z) & \Delta \hat{P}_{gg}(z) \end{matrix}\right) \left(\begin{matrix} L_q(x/z) \\ L_g(x/z) \\ \Delta \Sigma (x/z) \\ \Delta G(x/z) \end{matrix}\right)\,, \label{d2}
\eeq
where
\beq
&&\hat{P}_{qq}(z)=C_F \left(\frac{z(1+z^2)}{(1-z)_+} + \frac{3}{2}\delta(1-z) \right)\,, \\
&&\hat{P}_{qg}(z) = n_f z(z^2+(1-z)^2)\,, \\
&&\hat{P}_{gq}(z)= C_F(1+(1-z)^2)\,, \\
&&\hat{P}_{gg}(z)= 6\frac{(z^2-z+1)^2}{(1-z)_+} +  \frac{\beta_0}{2}\delta(z-1)\,, \\
&&\Delta \hat{P}_{qq}(z)=C_F (z^2-1)\,, \\
&&\Delta \hat{P}_{qg}(z)=n_f (1-z)(1-2z+2z^2)\,, \\
&&\Delta \hat{P}_{gq}(z)= C_F (z-1)(-z+2)\,, \\
&&\Delta \hat{P}_{gg}(z)=6 (z-1)(z^2-z+2)\,.
\eeq
Integrating both sides of (\ref{d2}) over $x$ from 0 to 1, we obtain
\beq
\frac{d}{d\ln Q^2} \left(\begin{matrix} L_q \\ L_g \end{matrix}\right)= \frac{\alpha_s}{2\pi}  \left(\begin{matrix} -\frac{4}{3}C_F  & \frac{n_f}{3}  & -\frac{2}{3}C_F & \frac{n_f}{3} \\  \frac{4}{3}C_F & -\frac{n_f}{3} & -\frac{5}{6}C_F & -\frac
{11}{2} \end{matrix}\right) \left(\begin{matrix} L_q\\ L_g \\ \Delta \Sigma  \\ \Delta G \end{matrix}\right)\,, \label{dd}
\eeq
in agreement with \cite{Ji:1995cu}. 
One can check that\footnote{ In fact, $\frac{d}{d\ln Q^2}\Delta \Sigma (Q^2)=0$ to this order.}
\beq
\frac{d}{d\ln Q^2} \left( \frac{1}{2}\Delta \Sigma (Q^2) + \Delta G (Q^2)+L_q(Q^2)+L_g(Q^2) \right)=0,
\eeq
that is, the total angular momentum $\frac{1}{2}$ is conserved.


\section{Numerical results}

In this section we show the result of our numerical simulation of the coupled equations (\ref{d1}) and (\ref{d2}) directly in the $x$-space.
We set $n_f=3$ throughout and use the one-loop running coupling constant $\alpha_s(Q^2)=\frac{4\pi}{\beta_0 \ln Q^2/\Lambda^2}$ with $\Lambda=0.25$ GeV. The initial scale is chosen to be $Q_0^2=1$ GeV$^2$. As for the initial condition, in principle one should draw guidance from the existing global analysis fits \cite{Hirai:2006sr,Leader:2010rb,deFlorian:2014yva,Nocera:2014gqa,Sato:2016tuz,Salajegheh:2018luz}. However, this is not practical for the present purpose due to several reasons. Firstly, since our primary interest is in the small-$x$ region, the details of parameterization in the large-$x$ region are presumably not very important. Besides, currently the uncertainties of $\Delta G(x)$ in the small-$x$ region are enormous. Moreover, there has been no global analysis study of the orbital angular momentum distribution. (There are, however, some model calculations \cite{Lorce:2011ni,Kanazawa:2014nha,More:2017zqp} and an estimate in the Wandzura-Wilczek approximation \cite{Rajan:2016tlg}.) 
We therefore restrict ourselves to the following very simple models:

\begin{itemize}

\item Democratic model \\
The nucleon spin is equally distributed to the four terms in (\ref{1}) at the initial scale. The distributions are nonsingular as $x\to 0$. For simplicity, we assume they are constant in $x$. 
\beq
\Delta\Sigma(x,Q_0^2) =  \frac{1}{4}, \quad \Delta G(x,Q_0^2)=\frac{1}{8}, \quad L_q(x,Q^2_0)=\frac{1}{8}, \quad L_g(x,Q_0^2)=\frac{1}{8},
\eeq

\item Helicity dominance model \\
Initially the helicity contributions alone saturate the sum rule. We try
\beq
\Delta\Sigma(x,Q_0^2) =A_q x^{-0.3}(1-x)^{3}, &\quad&  \Delta G(x,Q_0^2)=A_g x^{a_g}(1-x)^{3}, \nn
 L_q(x,Q^2_0)&=&L_g(x,Q_0^2)=0,  \label{initial}
\eeq 
 where   $A_q$ and $A_g$ are fixed by the conditions $\Delta \Sigma(Q_0^2)=\frac{1}{4}$ and $\Delta G(Q_0^2)=\frac{3}{8}$. 
We vary the parameter $a_g$ to explore different possibilities.\footnote{Refs.~\cite{deFlorian:2014yva,Salajegheh:2018luz} found a positive and large value $a_g\approx 1\sim 1.6$, whereas Ref.~\cite{Sato:2016tuz} found a slightly negative value $a_g\approx -0.15$. The uncertainties to these values are quite large, on the order of a few hundred percent.} 

\end{itemize}

\begin{figure}[h]
\begin{center}
 \includegraphics[height=50mm]{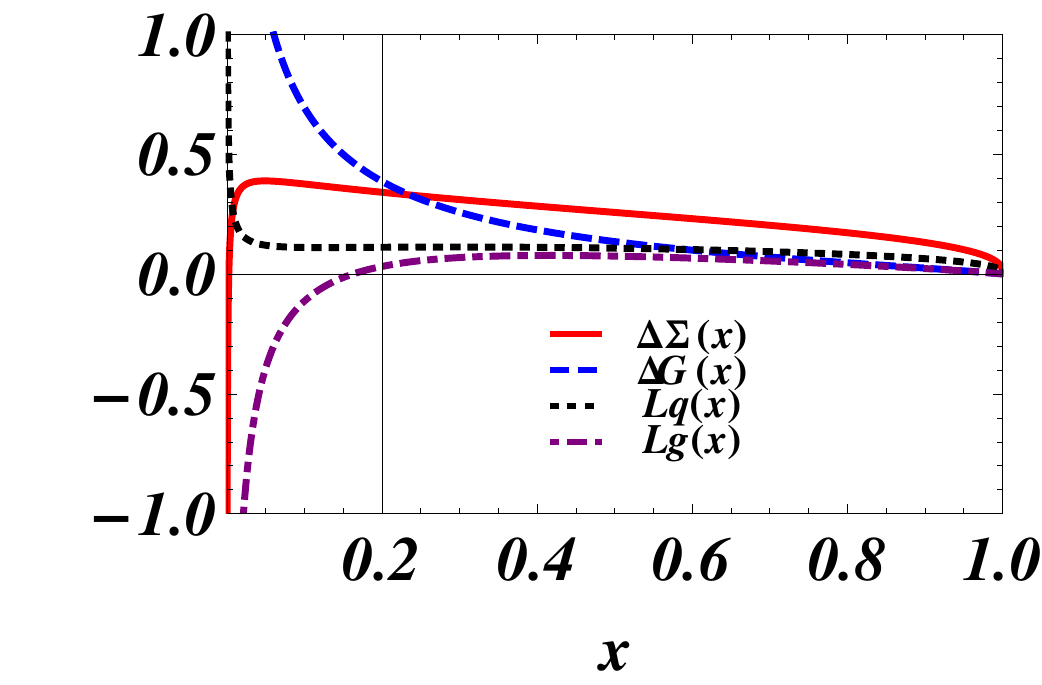}  
 \includegraphics[height=50mm]{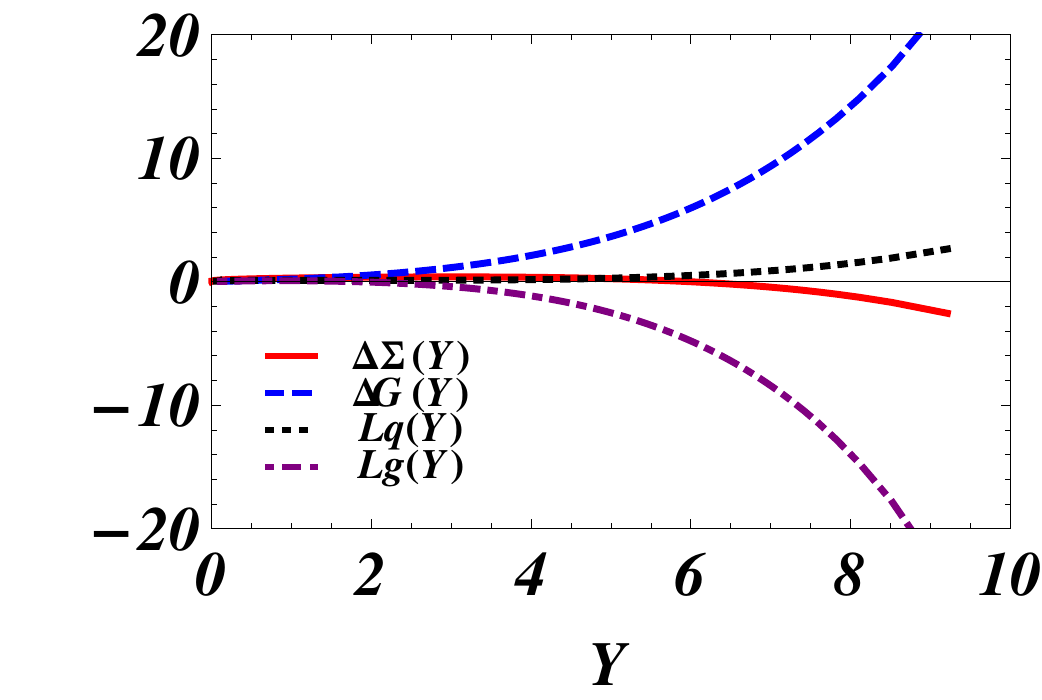}  
\end{center}
\caption{The $x$-distributions at $Q^2=10$ GeV$^2$ in the democratic model. In the small-$x$ region it is more convenient to use the rapidity variable $Y=\ln 1/x$. } 
\label{fig1}
\end{figure}

\begin{figure}[h]
 \includegraphics[height=51mm]{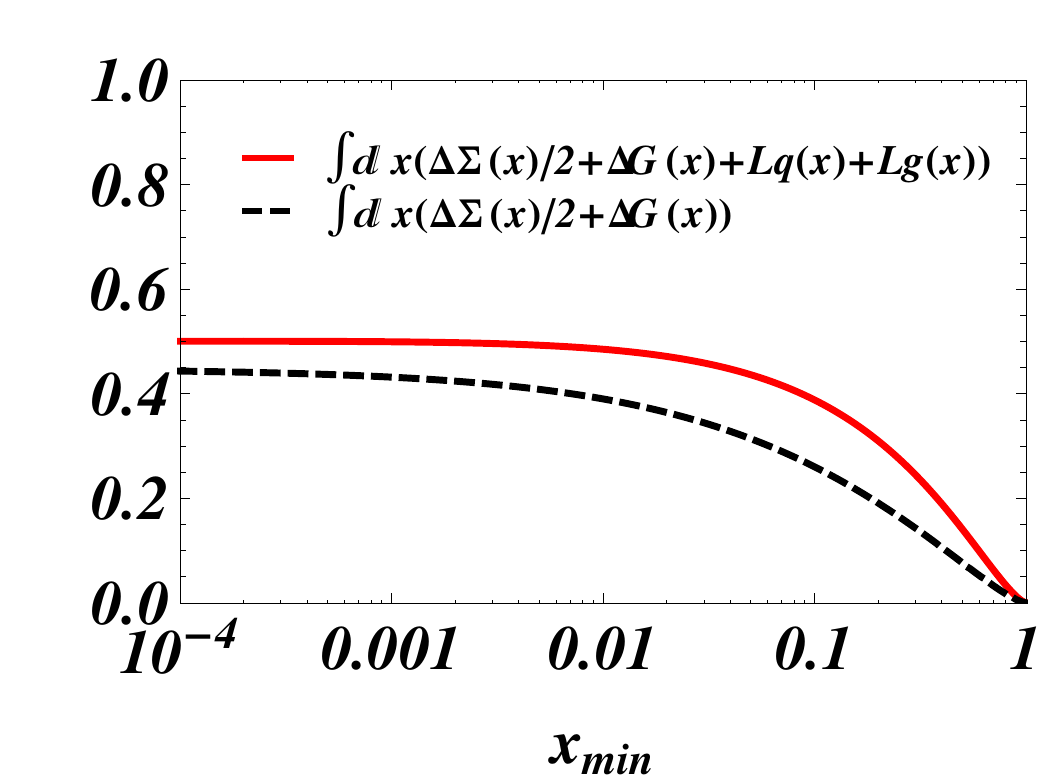} 
\includegraphics[height=51mm]{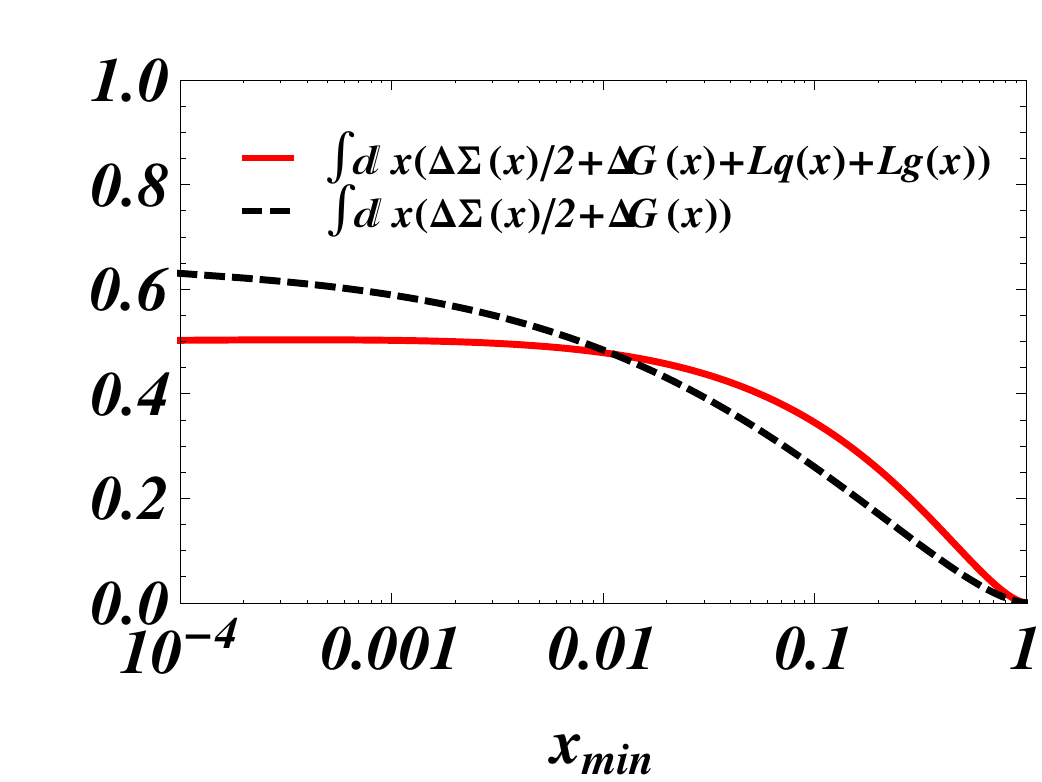} 
\caption{Solid line: The integral (\ref{min}) as a function of $x_{min}$ at $Q^2=10$ GeV$^2$ (left) and $Q^2=100$ GeV$^2$ (right). The dashed line includes only  the quark and gluon helicity contributions. } 
\label{fig2}
\end{figure}

Fig.~\ref{fig1}  shows $\Delta \Sigma(x),\Delta G(x),L_q(x),L_g(x)$ in the democratic model as a function of $x$ (left) and rapidity $Y\equiv \ln 1/x$ (right) at $Q^2=10$ GeV$^2$.   We find that at small-$x$, $\Delta \Sigma (x)$ and $L_g(x)$ turn negative and significantly cancel $L_q(x)$ and $\Delta G(x)$, respectively.  In Fig.~\ref{fig2}, we plot the  quantity
\beq
\int_{x_{min}}^1 dx \left(\frac{1}{2}\Delta\Sigma(x) + \Delta G(x)+ L_q(x)+L_g(x) \right),  \label{min}
\eeq
as a function of $x_{min}$ at $Q^2=10$ GeV$^2$ (left) and $Q^2=100$ GeV$^2$ (right). For comparison, we also plot the helicity part alone    $\int_{x_{min}}^1 dx \left(\frac{1}{2}\Delta\Sigma(x) + \Delta G(x)\right)$ (dashed line).        Depending on the value of $Q^2$, the helicity contribution undershoots (small-$Q^2$) or overshoots (large-$Q^2$) the total spin $1/2$. In either case, in this model the helicity contribution becomes the dominant part of the total spin, although initially it has the same magnitude as the orbital angular momentum contribution.

\begin{figure}[h]
 \includegraphics[height=48mm]{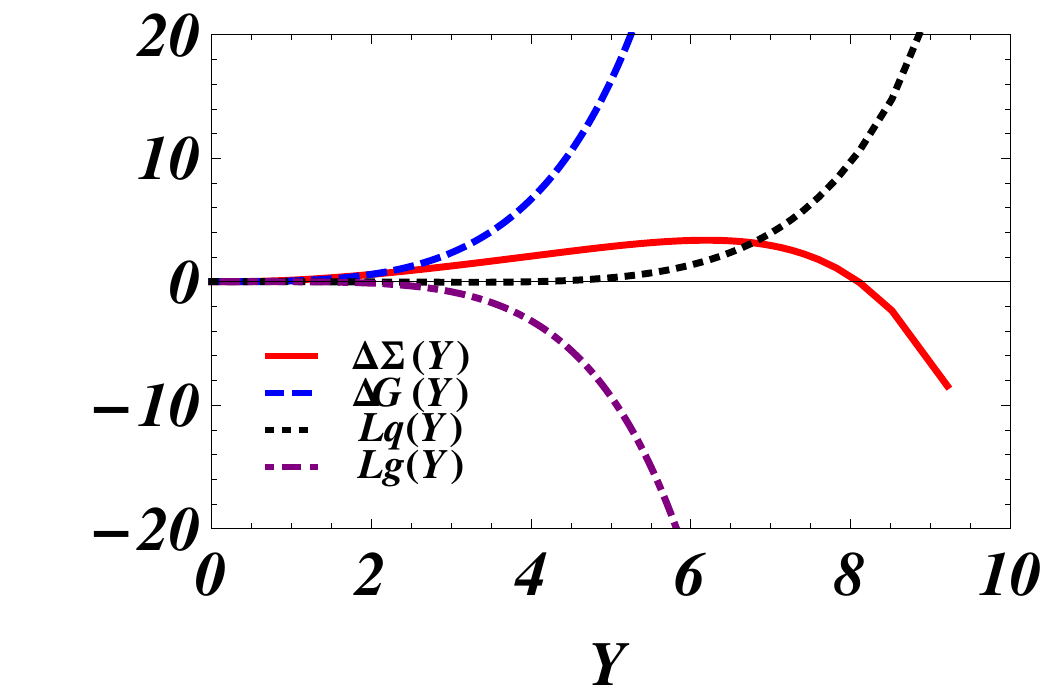} 
 \includegraphics[height=48mm]{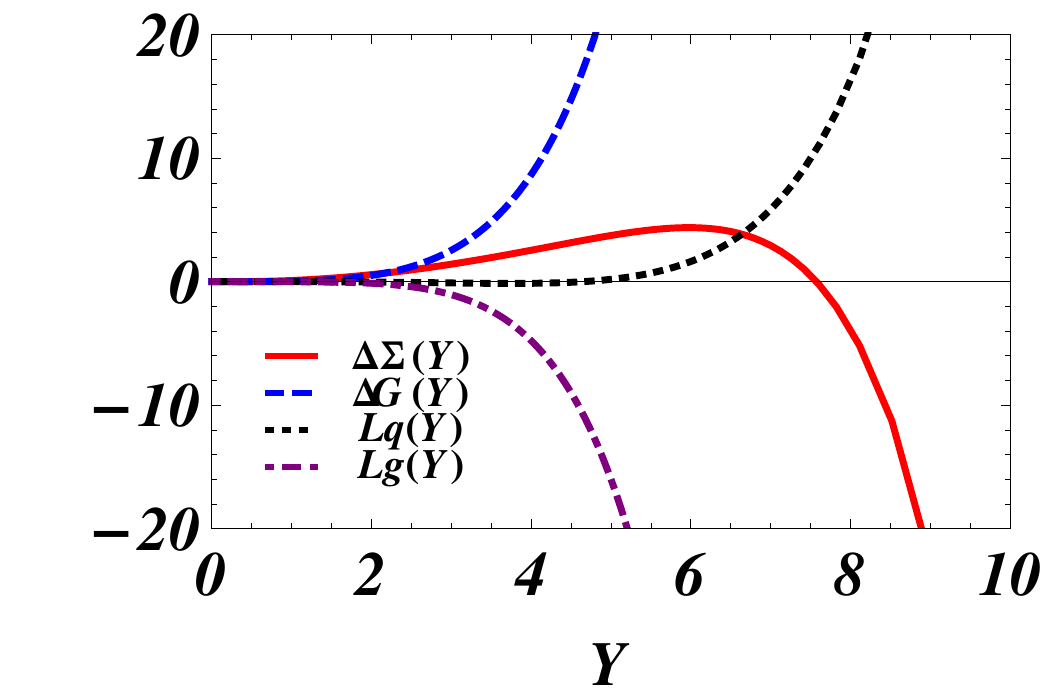} 
\caption{The $Y=\ln1/x$ distribution in the helicity dominance model with $a_g=-0.6$ at $Q^2=10$ GeV$^2$ (left) and at $Q^2=100$ GeV$^2$ (right).  } 
\label{fig3}
\end{figure}

\begin{figure}[h]
 \includegraphics[height=51mm]{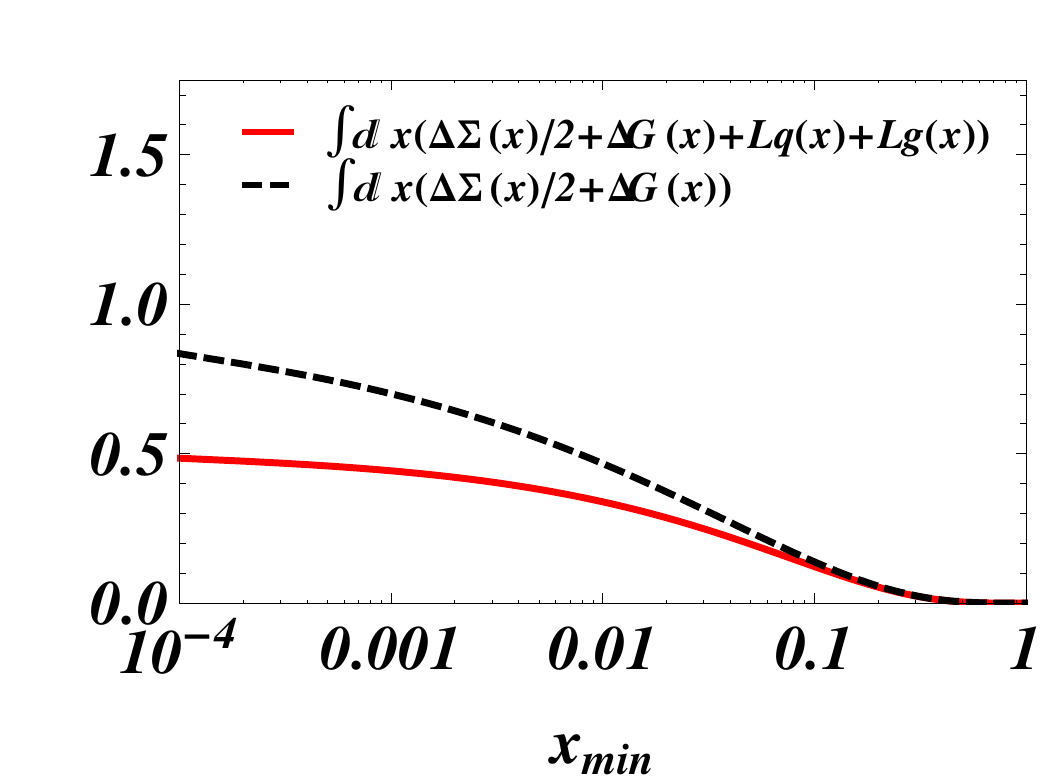} 
\includegraphics[height=51mm]{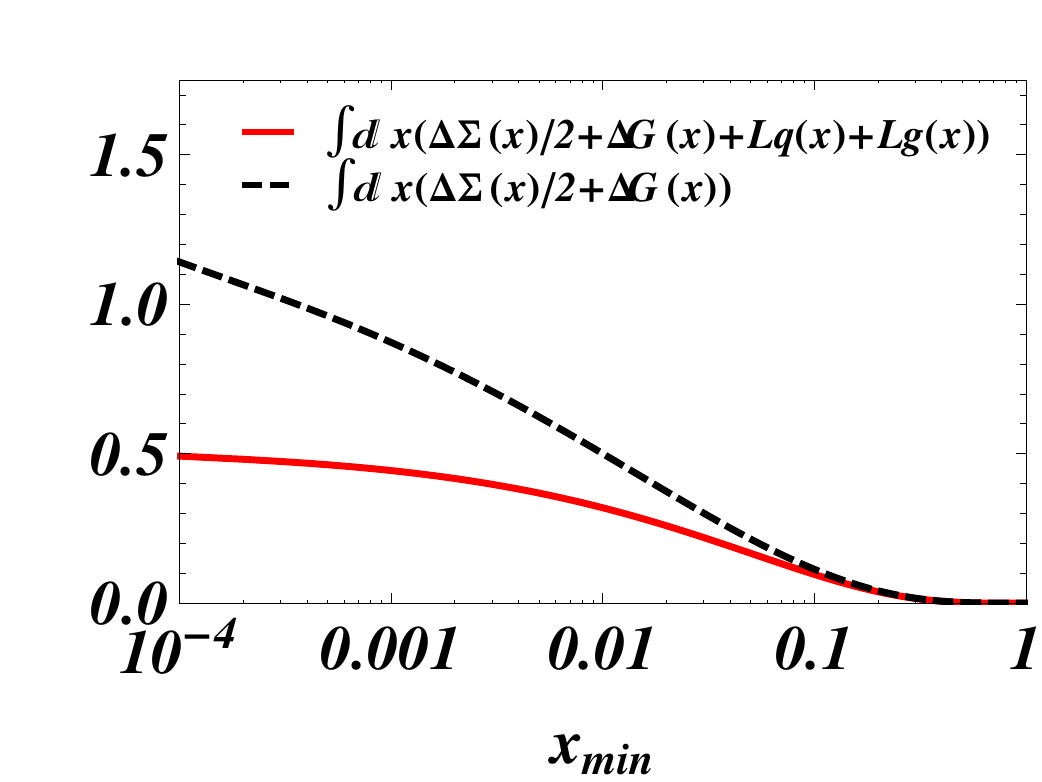} 
\caption{Solid line: The integral (\ref{min}) as a function of $x_{min}$ at $Q^2=10$ GeV$^2$ (left) and $Q^2=100$ GeV$^2$ (right) in the helicity dominance model. The dashed line includes only  the helicity contributions. } 
\label{fig4}
\end{figure}

Fig.~\ref{fig3} shows the result for the helicity dominance model with $a_g=-0.6$ at   $Q^2=10$ GeV$^2$ (left) and $Q^2=100$ GeV$^2$ (right). Although $\Delta \Sigma(x)$ is initially positive and has a power-law divergence, after the evolution again it turns negative and partly cancels $L_q(x)$.  $L_g(x)$ is initially zero, but it quickly develops a strong singularity and significantly cancels $\Delta G(x)$.  Fig.~\ref{fig4} is the plot of (\ref{min}) in this model. We see that even though the orbital angular momentum vanishes at $Q^2=1$ GeV$^2$, already at $Q^2=10$ GeV$^2$ it plays a crucial role to fulfill the spin sum rule. We also notice that  very little orbital angular momentum is generated in the large-$x$ region, $x>0.1$. 
 The result for $a_g=-0.3$ is qualitatively similar.  
$\Delta \Sigma(x)$ turns negative even  though initially it is as strongly divergent as $\Delta G(x)$. 

\begin{figure}[h]
\includegraphics[height=51mm]{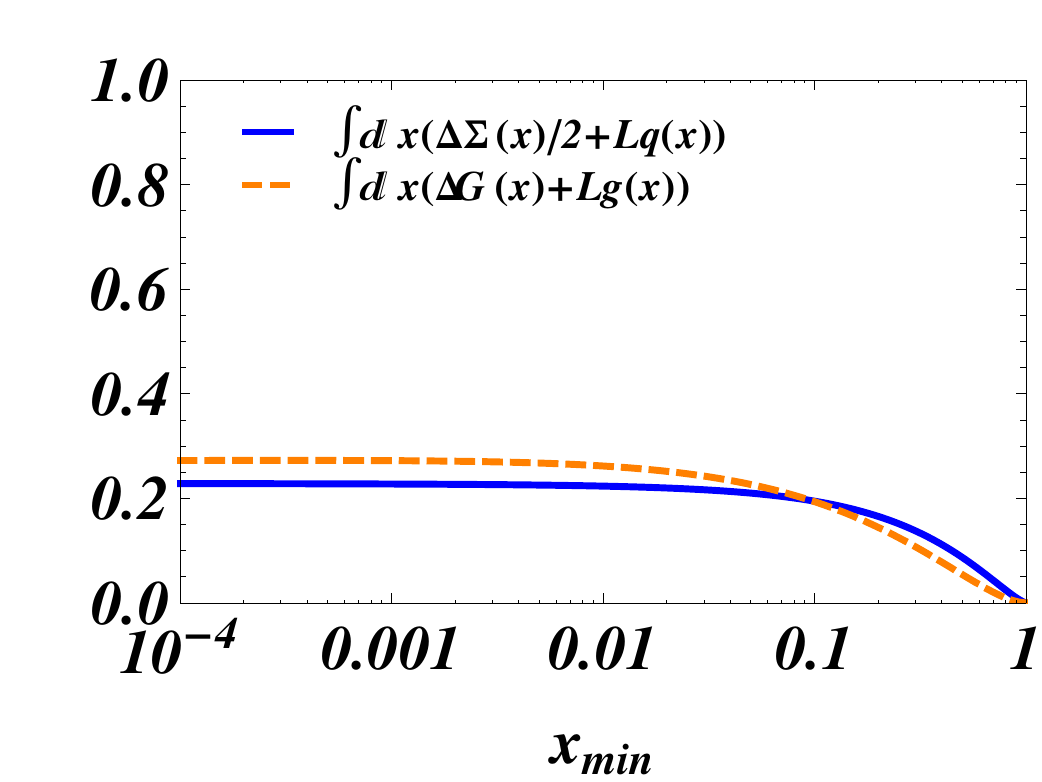} 
 \includegraphics[height=51mm]{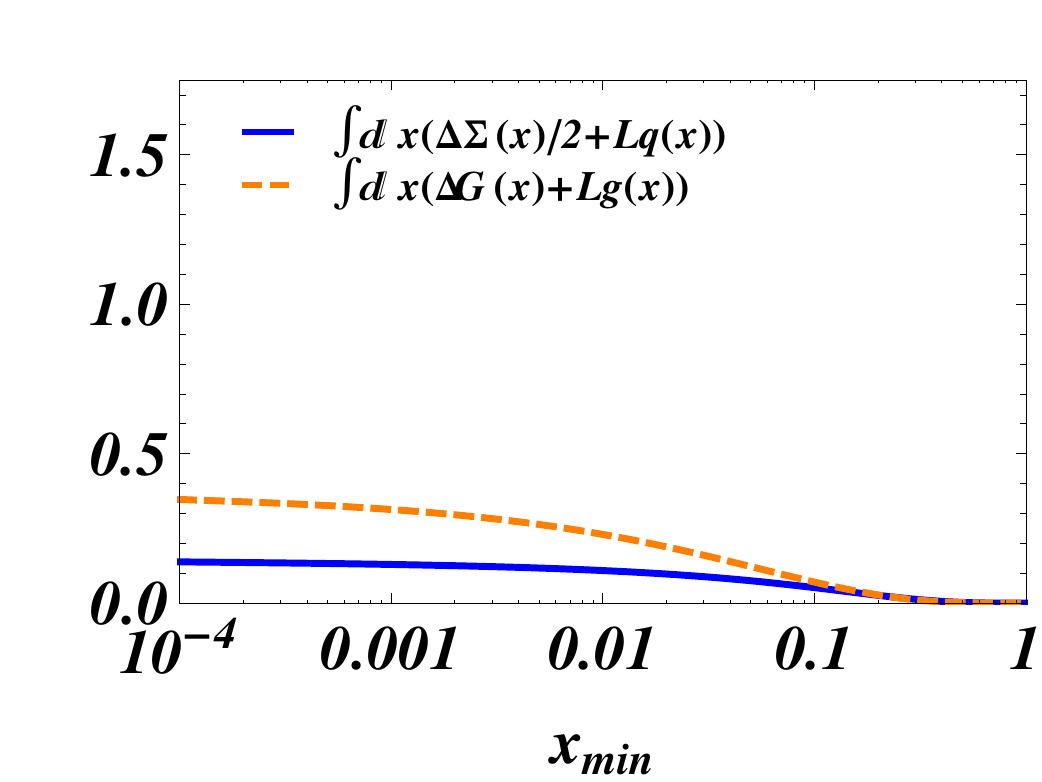} 
\caption{The quark (solid line) and gluon (dashed line) contributions to the nucleon spin in the democratic model (left) and the helicity dominance model (right) at $Q^2=10$ GeV$^2$. } 
\label{sep}
\end{figure}

Comparing the two models, we see that the growth of orbital angular momentum at small-$x$ is mainly governed by the parameter $a_g$, rather than the initial value $L_g(Q^2_0)$. If the initial helicity distribution is singular $a_g<0$, a significant amount of orbital angular momentum is generated at small-$x$. It is thus important to better constrain the value $a_g$ in future global analyses.   
Finally, in Fig.~\ref{sep} we plot the quark and gluon contributions to the nucleon spin
\beq
\int_{x_{min}}^1 dx \left(\frac{1}{2}\Delta\Sigma(x) +  L_q(x) \right), \qquad  \int_{x_{min}}^1 dx \left(\Delta G(x)+  L_q(x) \right),
\eeq
 separately in the two models at $Q^2=10$ GeV$^2$. Due to the cancellation, the two curves flatten quickly in the democratic model. In the helicity dominance model, a gradual rise of the gluon angular momentum is observed down to $x\sim 10^{-3}$,  implying that the cancellation is not complete.

\section{Analytical insights}

The significant cancellation between $\Delta G(x)$ and $L_g(x)$ and also between $\Delta\Sigma(x)$ and $L_q(x)$ we found numerically in the previous section is phenomenologically important and calls for a theoretical explanation. 
Let us try to understand by analytical means how such a cancellation can arise from the structure of the evolution equation. 

Since $|L_g(x)|, |\Delta G(x)| \gg |\Delta \Sigma(x)|, |L_q(x)|$ at small-$x$, to first approximation we may ignore $\Delta\Sigma(x)$ and $L_q(x)$ altogether. The equation then reads 
\beq
&&\frac{d}{d\ln Q^2} \left(\begin{matrix}  L_g(x) \\  \Delta G(x)  \end{matrix}\right) \nn 
&& \approx  \frac{\alpha_s}{2\pi} \int_x^1 \frac{dz}{z} \left(\begin{matrix}  6\frac{(z^2-z+1)^2}{(1-z)_+} +  \frac{\beta_0}{2}\delta(z-1)  & 6 (z-1)(z^2-z+2)  \\ 0 &6 \left(\frac{1}{(1-z)_+}-2z+1\right) +  \frac{\beta_0}{2}\delta(z-1) \end{matrix}\right) \left(\begin{matrix}   L_g(x/z)  \\ \Delta G(x/z) \end{matrix}\right). \label{red}
\eeq
 Let us first consider the double logarithmic approximation (DLA) which is familiar in the context of the unpolarized distributions but can be readily generalized to the helicity distributions \cite{Ahmed:1975tj,Einhorn:1985dy,Ball:1995ye}. In this approximation, (\ref{red}) reads, in the Mellin space,
\beq
\frac{d}{d\ln Q^2} \left(\begin{matrix}  L_g(j) \\   \Delta G(j)  \end{matrix}\right)\approx  \frac{\alpha_s}{2\pi} \left(\begin{matrix}  \frac{6}{j}-\frac{11}{2}-\frac{n_f}{3} &   \frac{-12}{j}+14   \\
0  & \frac{12}{j} -\frac{13}{2}-\frac{n_f}{3} \end{matrix}\right) \left(\begin{matrix} L_g(j)  \\ \Delta G(j) \end{matrix}\right)\,, \label{back}
\eeq
where we expanded around the singularity at $j=0$. 
(It is straightforward to include $\Delta \Sigma(x)$ and $L_q(x)$, see Appendix.)   
Diagonalizing the matrix and following the standard procedure,  one finds
\beq
\Delta G(x,Q^2) &\approx& \frac{\xi^{\frac{1}{4} }}{\sqrt{\pi} (2Y)^{3/4}} e^{2\sqrt{2\xi Y}-\xi\left(\frac{13}{12}+\frac{n_f}{18}\right)} \Delta G(j_0,Q_0^2), \label{dla} \\
L_g(x,Q^2) + 2\Delta G(x,Q^2)&\approx& \frac{\xi^{\frac{1}{4} }}{2\sqrt{\pi} Y^{3/4}} e^{2\sqrt{\xi Y}-\xi\left(\frac{11}{12}+\frac{n_f}{18}\right)} (L_g(j_0/\sqrt{2},Q_0^2)+2\Delta G(j_0/\sqrt{2},Q_0^2)),
\label{dla2}
\eeq
where $Y=\ln 1/x$, $\xi=\frac{12}{\beta_0} \ln \frac{\ln Q^2/\Lambda^2}{\ln Q_0^2/\Lambda^2}$ 
 and $j_0= \sqrt{2\xi/Y}$ is the saddle point of the inverse Mellin transform. In the asymptotic region $Y,\xi \to \infty$,  $|L_g(x)+2\Delta G(x)| \ll |\Delta G(x)|,|L_g(x)|$, or equivalently, $L_g(x) \approx -2\Delta G(x)$ \cite{Hatta:2016aoc}. 
However,  for realistic values of $Y,\xi$, the right hand side of (\ref{dla2}) is numerically not negligible. Besides, the exact eigenvector in the Mellins space is 
\beq
L_g(j)+\frac{2(6-7j)}{6-j}\Delta G(j),
\eeq
and to obtain (\ref{dla2}) we have approximated this as $L_g(j)+2\Delta G(j)$. 
These subleading effects tend to reduce the ratio $|L_g(x)/\Delta G(x)|$. Indeed numerically we find $|L_g(x)/\Delta G(x)|\approx 1$. 

\begin{figure}[h]
\begin{center}
 \includegraphics[height=61mm]{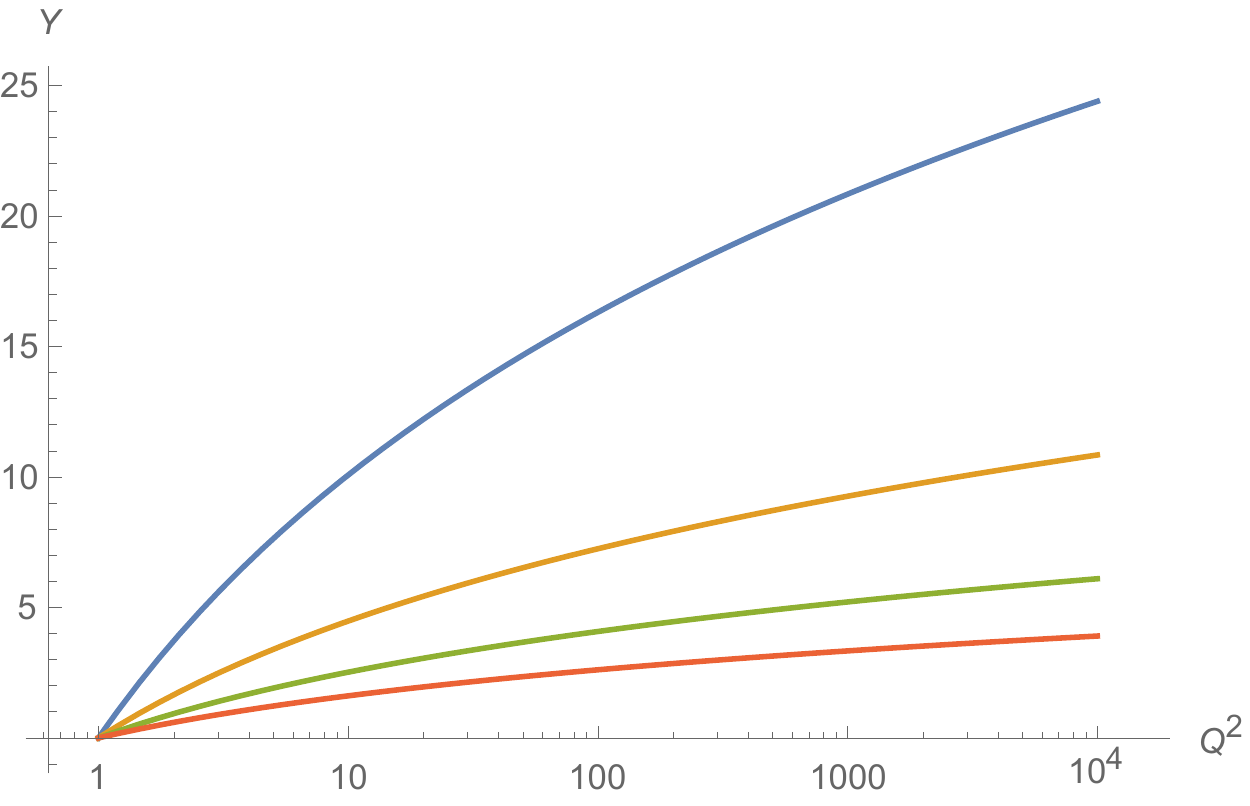} 
\end{center} 
\caption{The boundary between the DLA and power-law regimes for different values of $c=0.4,0.6,0.8,1$ (top to bottom) in the running coupling case. Above the boundary, the initial power-law survives. Below the boundary, the DLA is valid. } 
\label{figc}
\end{figure}

In deriving (\ref{dla}), it has been assumed that the DLA saddle point $j=j_0$ is to the right of all poles in the complex $j$-plane. This may not be  the case if the initial condition (\ref{initial}) has a singularity $a_g \equiv -c <0$.  When $j_0>c$, the DLA saddle point rules and the initial power-law is washed out. On the other hand, when $c>j_0$, an extra  term   
\beq
\sim \frac{e^{2\xi/c}}{x^c}, \label{ex}
\eeq
appears, and this will dominate over (\ref{dla}). 
The boundary $j_0=c$ forms a two-dimensional surface in the $(Y, \ln Q^2, c)$ space as illustrated in Fig.~\ref{figc}. 
A recent global analysis  \cite{Nocera:2014uea} found\footnote{More precisely, Ref.~\cite{Nocera:2014uea} computed the effective power
\beq
c\equiv \frac{\partial \ln |\Delta G(x)|}{\partial \ln 1/x}, 
\eeq
and found that the dependence of $c$ on $x$ is weak.} a rather large value  $1\gtrsim  c \gtrsim 0.5$  for $\Delta G(x)$ at $Q^2=4$ GeV$^2$ and $x<10^{-3}$.
This suggests that the power-law regime (\ref{ex}) indeed exists and is accessible in the present and future experiments. 

We now show that in the power-law regime, the ratio $L_g(x)/\Delta G(x)$ is directly related to the `Regge intercept' $c$. For this purpose we look for a solution whose leading singularity is of the form
\beq
L_g(x,Q^2) \approx A(Q^2)\frac{1}{x^c},  \qquad  \Delta G(x,Q^2)\approx B(Q^2)\frac{1}{x^c}.  \label{ans}
\eeq
 In order for the first moments to be finite, we assume $1 > c$. (The boundary value $c=1$ is actually interesting, see below.)
Substituting (\ref{ans}) into (\ref{red}), we can perform the $z$-integral analytically.\footnote{To be more precise, since (\ref{ans}) is valid only for $x\ll x_0$ with some $x_0 < 1$, one has to divide the $z$-integral into different regions $\int_x^1dz =\int_x^{x/x_0}dz + \int_{x/x_0}^1dz$.  It is easy to see that only the second integral, where the form (\ref{ans}) can be used, contributes to the leading singularity.} In practice, the lower limit of the integral can be sent to zero since the neglected terms are subleading in $x$. We  thus obtain  
\beq
\frac{d}{d\ln Q^2} \left(\begin{matrix} A(Q^2) \\  B(Q^2)  \end{matrix}\right)\approx  \frac{\alpha_s}{2\pi}  \left(\begin{matrix} 
 \alpha & \beta \\ 0 & \delta \end{matrix}\right) \left(\begin{matrix}   A(Q^2) \\ B(Q^2) \end{matrix}\right)\,, \label{dg}
\eeq
where 
\beq
\alpha &=& 6 \left( -H_{c-1}-\frac{2}{c+1}+\frac{1}{c+2} -\frac{1}{c+3}\right) + \frac{\beta_0}{2}, \\
\beta &=& 6\left( \frac{3}{c+1}-\frac{2}{c+2}+\frac{1}{c+3}-\frac{2}{c} \right),
 \\
\delta &=& 6 \left(  -H_{c-1}+\frac{1}{c}-\frac{2}{1+c} \right) + \frac{\beta_0}{2}.
\eeq
 ($H_x=x\sum_{k=1}^\infty \frac{1}{k(x+k)}$ is the harmonic number.) $\alpha,\beta,\delta$ are the anomalous dimensions \cite{Hagler:1998kg} analytically continued to noninteger values $j\to c$.\footnote{The entries in (\ref{back}) are the first two terms of $\alpha,\beta,\delta$ in the limit $c\to 0$.}   Eq.~(\ref{dg}) can be readily solved as  
\beq
B(Q^2) =C (\ln Q^2/\Lambda^2)^{\frac{2\delta}{\beta_0}}, \qquad A(Q^2) = C\frac{\beta }{\delta-\alpha}  (\ln Q^2/\Lambda^2)^{\frac{2\delta}{\beta_0}} + C'   (\ln Q^2/\Lambda^2)^{\frac{2\alpha}{\beta_0}} ,  \label{ab}
\eeq 
where $C,C'$ are the integration constants. 
Since $\delta >0$ and $\delta> \alpha$ for $1\ge c \ge 0$, the second term in $A$ is formally subleading (though it may not be  numerically subleading in practice). This gives 
\beq
\frac{L_g(x,Q^2) }{\Delta G(x,Q^2) }\approx  \frac{\beta}{\delta-\alpha} =-\frac{2}{c+1}, \label{c}
\eeq
 for sufficiently large $Q^2$.
We see that $L_g(x)$ and $\Delta G(x)$  have opposite signs at small-$x$, and that $L_g(x)$ is larger in magnitude than $\Delta G(x)$. 
The DLA corresponds to the limit $c\to 0$ where the anomalous dimensions diverge and the distributions depend logarithmically on $x$ (instead of a power-law).

\begin{figure}[h]
\begin{center}
 \includegraphics[height=61mm]{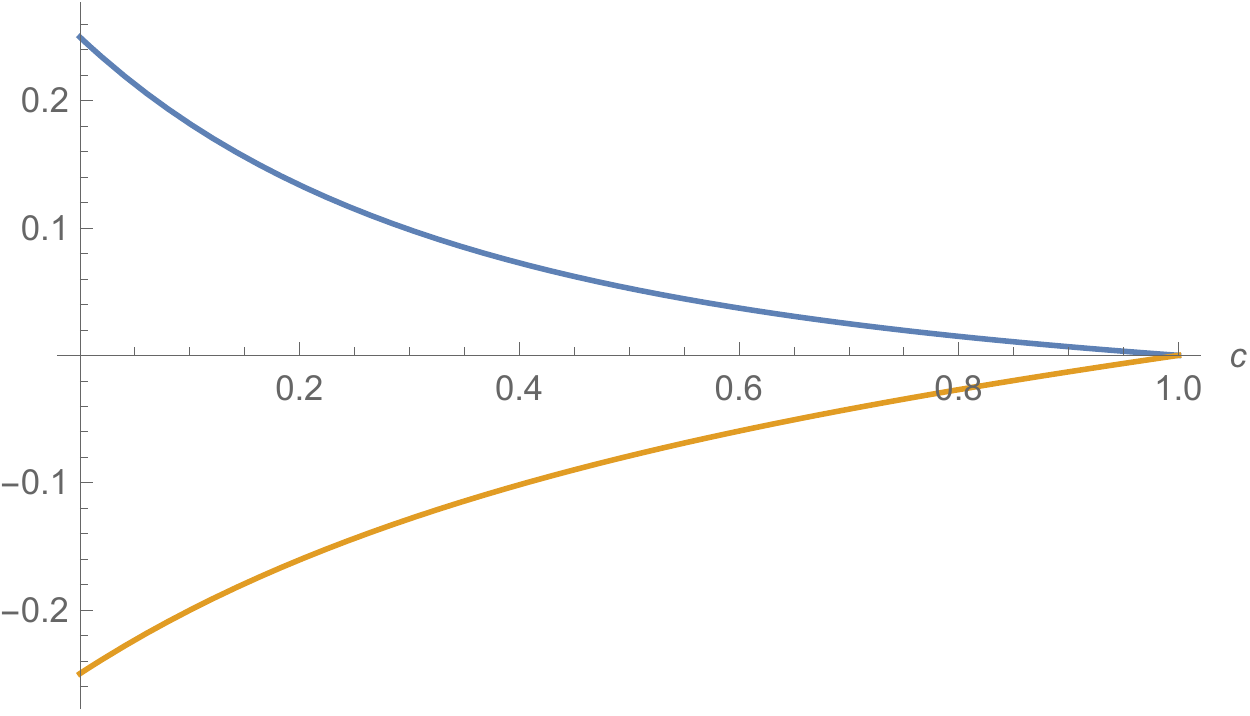} 
\end{center} 
\caption{Lower curve: $\Delta \Sigma(x)/\Delta G(x)$ from (\ref{ratio}) as a function of $c$ with $n_f=3$; Upper curve: $L_q(x)/\Delta G(x)$ from (\ref{ratio2}). } 
\label{ratiofig}
\end{figure}

Similarly, the small-$x$ behavior of $\Delta\Sigma(x)$ is governed by the exponent $c$. Keeping only $\Delta G(x)$ from (\ref{ans}) on the right hand side of (\ref{d1}), one finds
\beq
\frac{d}{d\ln Q^2}\Delta \Sigma(x) \approx \frac{n_f\alpha_s}{2\pi} \int_x^1\frac{dz}{z} (2z-1)\Delta G(x/z) \approx -\frac{n_f \alpha_s}{2\pi}\frac{1-c}{ c(1+c)} \Delta G(x).  
\eeq
This immediately gives
\beq
\frac{\Delta\Sigma(x)}{\Delta G(x)} \approx -n_f \frac{1-c}{\delta c(1+c)} = -n_f \frac{1-c}{c(1+c)\left[6\left(-H_{c-1}+\frac{1}{c}-\frac{1}{1+c}\right)+\frac{\beta_0}{2}\right]}.  \label{ratio}
\eeq
The ratio is plotted in Fig.~\ref{ratiofig} (lower curve). $\Delta \Sigma(x)$ has an opposite sign to $\Delta G(x)$ and its magnitude is strongly suppressed as $c\to 1$. This is consistent with the numerical results in the previous section. 
Finally, for the quark orbital angular momentum, we find
\beq
\frac{d}{d\ln Q^2}L_q(x) &\approx& \frac{n_f\alpha_s}{2\pi}\int_x^1\frac{dz}{z} \left( z(z^2+(1-z)^2)L_g(x/z) + (1-z)(1-2z+2z^2)\Delta G(x/z) \right)\nn 
&\approx& \frac{n_f\alpha_s}{2\pi} \frac{1-c}{c(1+c)^2}\Delta G(x), 
\eeq
so that
\beq
L_q(x) \approx  n_f \frac{1-c}{\delta c(1+c)^2} \Delta G(x) \approx  -\frac{\Delta \Sigma(x)}{1+c}. \label{ratio2}
\eeq
This is also plotted in Fig.~\ref{ratiofig} (upper curve).

The limit $c\to 1$ is particularly interesting. In this limit, 
\beq
L_g(x) \approx -\Delta G(x), \quad L_q(x)\approx -\frac{1}{2}\Delta \Sigma(x).  \label{ag}
\eeq
We see that the boundary value $c=1$ is permissible from the sum rule point of view. While $\int \Delta G(x)$ and $\int L_g(x)$ are both logarithmically divergent, the divergent parts cancel exactly. Interestingly, the first relation in  (\ref{ag}) formally  agrees with the argument in \cite{Hatta:2016aoc} based on  an operator analysis without any reference to the small-$x$ behavior of $\Delta G(x),L_g(x)$.\footnote{In \cite{Hatta:2016aoc}, there was a mistake in the normalization of $L_g(x)$ from Section III on. Because of this,  the correct relation $L_g(x) \approx -\Delta G(x)$ was incorrectly presented as  $L_g(x) \approx -2\Delta G(x)$ and this  coincided with the DLA result in the appendix of \cite{Hatta:2016aoc} (where the normalization is correct), creating an apparent `consistency.'  } 
It has been also observed in an explicit model calculation in \cite{More:2017zqp}.

In practice, it is difficult to numerically confirm the ratios and exponents obtained above. This is because the approach to the asymptotic regime is slow, especially due to the running of the coupling. For realistic values of $Q^2$ and $x$,  the subleading corrections which depend on the initial condition are still not negligible.  For example, in Fig.~\ref{fig3} we find $|\Delta G(x)| > |L_g(x)|$, although asymptotically the inequality should be reversed, see (\ref{c}). (Note that in this model  $L_g(x)=0$ initially, so the second term of $A$ in (\ref{ab}) can be comparable to the first term.) Also,  whether the $Y$-dependence is exponential $e^{\# Y}$ or DLA-like $e^{\# \sqrt{Y}}$, is difficult to determine because of the limited $Y$-range. We however checked that as $c=-a_g$ is increased, a DLA-like fit becomes less and less favorable.\footnote{For example, at $Q^2=10$ GeV$^2$, DLA predicts $\Delta G(x) \sim e^{2.53\sqrt{Y}}$, while our fit  is $\sim e^{3.06\sqrt{Y}}$ for $a_g=-0.3$ and $\sim e^{3.96\sqrt{Y}}$ for $a_g=-0.6$ in the helicity dominance model. The exponential fit instead gives $e^{0.55Y}$ for $a_g=-0.3$ and $e^{0.70Y}$ for $a_g=-0.6$. }


The  picture emerging from our analysis is that, for any value of $c$ between 0 and 1, among the four terms in the Jaffe-Manohar decomposition in the $x$-space,
\beq
\frac{1}{2}\Delta\Sigma (x)+ L_q(x) + \Delta G(x)+L_g(x),
\eeq
there is a significant cancellation between the first two terms (quark sector) and also between the last two terms (gluon sector). 
Note that when $c=1$ exactly, in practice one is  computing the anomalous dimensions relevant to the first moments (\ref{int}), see (\ref{dd}). Therefore, a similar cancellation occurs among the integrated quantities $\frac{1}{2}\Delta \Sigma+L_q$ and $\Delta G+L_g$ as was observed in \cite{Ji:1995cu,Stratmann:2007hp}. We have shown that such a cancellation already occurs in the density space at small-$x$, and the deviation from exact cancellation is controlled by the Regge intercept $c$. 

\section{Conclusions}

In this paper we numerically solved the QCD evolution equation for the orbital angular momentum distributions. Compared with the previous work 
\cite{Martin:1998fe}, our work is focused on the small-$x$ region where an interesting cancellation occurs between $L_g(x)$ and $\Delta G(x)$ and also between $\Delta\Sigma(x)$ and $L_q(x)$. For $\Delta G(x)$, such a cancellation has been previously suggested in \cite{Hatta:2016aoc,More:2017zqp} from different arguments. As we demonstrated analytically, in the present approach this naturally follows from the structure of  the evolution equation. 

Our finding has an important  implications for phenomenology.  On one hand, the precise value of $\Delta G$ is of intrinsic interest in QCD, and it is certainly imperative to reduce the uncertainties of $\Delta G(x)$ in the small-$x$ region in future experiments such as at the planned Electron-Ion Collider (EIC). On the other hand, this is not sufficient to solve the nucleon spin puzzle because a good fraction of the would-be spin from $\Delta G(x)$ at small-$x$ is canceled by the orbital angular momentum in the same $x$-region. This suggests that one has to look into the orbital angular momentum in the large-$x$ region \cite{Ji:2016jgn,Bhattacharya:2017bvs}. After all, this is a very natural and obvious future direction of research.    

As already mentioned in the introduction, the DGLAP-type evolution equation considered in this paper eventually breaks down and should be superseded by the small-$x$ evolution equation which resums double logarithmic contributions $(\alpha_s \ln^2 1/x)^n$. (Not to be confused with the DLA which  resums powers of $\alpha_s \ln 1/x \ln Q^2$.)  For the helicity distributions, it is known that such a resummation dynamically generates a power-law behavior 
 \cite{Bartels:1996wc,Bartels:1995iu,Kovchegov:2015pbl,Kovchegov:2016zex,Kovchegov:2017lsr}. Furthermore, there may be a regime where nonlinear  evolution equations come into play, as is the case for the unpolarized distributions. Unfortunately, at the moment very little is known about the small-$x$ resummation for the orbital angular momentum distributions.
Admittedly, we may have pushed our numerical solution to too small values of $x$ for which the present approach is not  justified and an alternative approach is needed. Still, one can naturally expect that the rapid growth of the distributions, either due to the DLA  or the Regge behavior augmented by the QCD evolution, is smoothly connected with the power-law generated by the small-$x$ resummation.   Clearly this issue deserves further study.

\section*{Acknowledgments}
Y.~H. thanks Matthias Burkardt,  Cedric Lorce and Barbara Pasquini for a discussion concerning Footnote 7. We thank Shunzo Kumano, Nobuo Sato and Werner Vogelsang for  useful correspondence. 
D.~Y. is supported by the Ministry of Science and Technology (MOST) 
Grant No. 106-2112-M-033-003, 105-2112-M-033-004, and 107-2917-I-003-004.

\appendix
\section{DLA in the presence of quark distributions}

It is straightforward to generalize (\ref{back}) by including the quark distributions $\Delta \Sigma(x)$ and $L_q(x)$. 
\beq
\frac{d}{d\ln Q^2} \left(\begin{matrix} L_q(x) \\ L_g(x) \\  \Delta \Sigma(x) \\ \Delta G(x)  \end{matrix}\right)\approx  \frac{\alpha_s}{2\pi} \int_x^1 \frac{dz}{z} \left(\begin{matrix} 0 & 0  & -C_F & n_f \\ 2C_F & 6 & -2C_F &  -12 \\
0 & 0 & C_F & -n_f \\ 
0 & 0 & 2C_F & 12 \end{matrix}\right) \left(\begin{matrix} L_q(x/z) \\ L_g(x/z) \\ \Delta \Sigma (x/z) \\ \Delta G(x/z) \end{matrix}\right). 
\eeq
The following linear combinations diagonalize the evolution equation
\beq
S_1(x)&=&\frac{4}{9}(L_q(x)+\Delta\Sigma(x)) + L_g(x) + 2\Delta G(x), \\
S_2(x)&=&L_q(x)+\Delta \Sigma(x), \\
S_3(x)&=&\frac{1}{\sqrt{16-\frac{3n_f}{2}}} \Delta \Sigma(x) + \left(\frac{2}{\sqrt{16 -\frac{3n_f}{2}}} -\frac{1}{2}\right)\Delta G(x)\\
    &\approx& 0.3 \Delta \Sigma(x) +0.1\Delta G(x),  \qquad (n_f=3) \\
S_4(x) &=& \frac{1}{\sqrt{16-\frac{3n_f}{2}}} \Delta \Sigma(x) + \left(\frac{2}{\sqrt{16 -\frac{3n_f}{2}}}+\frac{1}{2}\right) \Delta G(x)
\nn 
&\approx & 0.3\Delta\Sigma(x) + 1.1\Delta G(x), \qquad (n_f=3)
\eeq
 such that
\beq
\frac{d}{d\ln Q^2} \left(\begin{matrix} S_1(x) \\ S_2(x) \\  S_3(x) \\ S_4(x)  \end{matrix}\right) &\approx&  \frac{\alpha_s}{2\pi} \int_x^1 \frac{dz}{z} \left(\begin{matrix} 6 & 0  &0& 0 \\ 0 & 0  & 0 &  0 \\
0 & 0 &  \frac{20}{3}-\frac{4}{3}\sqrt{16-\frac{3n_f}{2}}  & 0 \\ 
0 & 0 & 0 & \frac{20}{3}+\frac{4}{3}\sqrt{16-\frac{3n_f}{2}} \end{matrix}\right) \left(\begin{matrix} S_1(x/z) \\ S_2(x/z) \\ S_3(x/z) \\ S_4(x/z) \end{matrix}\right) \nn 
&\approx &  \frac{\alpha_s}{2\pi} \int_x^1 \frac{dz}{z} \left(\begin{matrix} 6 & 0  &0& 0 \\ 0 & 0  & 0 &  0 \\
0 & 0 & 2.1  & 0 \\ 
0 & 0 & 0 &11.2 \end{matrix}\right) \left(\begin{matrix} S_1(x/z) \\ S_2(x/z) \\ S_3(x/z) \\ S_4(x/z) \end{matrix}\right) .  \qquad (n_f=3)
\eeq
Asymptotically, $S_4(x)$ dominates and we have the relation
\beq
L_g(x) \approx -2\Delta G(x), \quad L_q(x) \approx -\Delta \Sigma(x) \approx \left(2-\sqrt{4-\frac{3n_f}{8}}\right)\Delta G(x) \approx 0.3\Delta G(x).
\eeq
This may be compared with the $c\to 0$ limit of (\ref{ratio}), $\Delta\Sigma (x) \approx -0.25 \Delta G(x)$ (for $n_f=3$). The difference is because  (\ref{ratio}) is obtained by first approximating $|\Delta \Sigma(x)|\ll |\Delta G(x)|$.

\end{document}